# EFFECTS OF MICROWAVE ABSORPTION ON LONG AND SHORT SINGLE-WALLED CARBON NANOTUBES AT $10^{-6}$ TORR

S. FERGUSON*[+], P. BHATNAGAR*, I. WRIGHT*, G. SESTRIC* and S. WILLIAMS*[#]

*Department of Physics and Geosciences
Angelo State University, ASU Station #10904
San Angelo, Texas 76909, USA

[+]Department of Physics
University of Texas at Dallas, 800 West Campbell Rd.
Richardson, Texas 75080, USA

[#]scott.williams@angelo.edu

Carbon nanotubes have been observed to emit ultraviolet, visible, and infrared radiation when exposed to microwaves. We have performed experiments in which both short (0.5 μm - 2 μm) and long (5 μm - 30 μm) single and double-walled carbon nanotubes were exposed to 2.46 GHz microwaves at a pressure of ~$10^{-6}$ Torr. Structural modifications of the carbon nanotubes due to microwave absorption have been studied using the Raman spectroscopy G-band and D-band intensities, which suggest that microwave irradiation at relatively low pressure results in an increase in nanotube defects, especially in the case of the long nanotubes. Furthermore, a comparison of the spectra of the radiation emitted from the nanotubes suggests that the longer nanotubes emitted radiation of much greater intensity than the shorter nanotubes. Based on the results of the experiments and results described in previous reports, the ultraviolet, visible, and infrared radiation emitted as the result of microwave absorption by carbon nanotubes seems to be primarily blackbody radiation emitted due to Joule heating. However, the presence of several broad photopeaks in the spectra of the emitted radiation (which do not seem to be related to gases absorbed by the nanotubes or the presence of catalyst particles) suggest that emissions are not the result of Joule heating alone.

## 1. Introduction

Microwave radiation is commonly used in processes related to the functionalization[1,2] and purification[3] of carbon nanotubes (CNTs). The behavior of CNTs in microwave fields, however, is not completely understood. Imholt *et al.*[4] observed CNTs to emit infrared, visible, and ultraviolet radiation when exposed to microwaves. The same study[4] also noted that the CNTs were heated to over 2000 K when irradiated with microwaves. Since then, there have been several other reports of the same phenomenon[5,6,7] and considerable controversy concerning the mechanism responsible for the phenomenon. It has been suggested that the heating of metal catalyst particles[8], Joule heating[9], transverse parametric resonance as the result of mechanical vibrations[10], the effects of substrates used to synthesize CNTs[11], and field emission-induced luminescence[6,7] could all possibly be mechanisms responsible for the observed emissions of infrared, visible, and ultraviolet radiation from CNTs exposed to microwaves.

Microwave-assisted acid digestion is a common technique used to purify CNT samples. The technique involves acid absorbing energy from the microwaves and dissolving metal catalysts while CNT walls are left unaffected. The covalent functionalization of CNTs is also commonly achieved using microwaves. Microwave irradiation of CNTs has been shown to reduce reaction times and result in CNTs with higher degrees of functionalization than CNTs functionalized by conventional thermal methods[9]. Furthermore, it has recently been shown that CNTs can be functionalized in solvent-free conditions using microwaves in a manner that involves relatively short amounts of time and is suitable for large-scale functionalization[12]. Despite the fact that there have been many investigations related to the microwave-assisted purification and functionalization of CNTs, there have been surprisingly few experiments performed in which the behavior of CNTs in microwave fields has been studied without the involvement of chemical reagents. Other than the aforementioned study performed by Imholt et al.[4] and a study performed by Alvarez-Zauco et al.[5] involving the microwave irradiation of multi-walled CNTs at a pressure of 0.1 Torr, the authors of this report are unaware of any studies performed by other groups in which the behavior of CNTs exposed to microwave fields has been studied in the absence of chemical reagents and in a low-pressure environment.

A recent study by our group[6], in which the spectra of the radiation emitted by both single-walled and multi-walled CNTs at ~$10^{-6}$ Torr during several microwave-irradiation and cooling cycles were compared, showed that the intensity of radiation emitted by the CNTs increased each time the CNTs were irradiated with microwaves. This behavior was initially attributed to the opening of CNT ends as the samples were heated, resulting in a decrease in the work functions. However, the spectra of the radiation emitted by the CNTs during microwave irradiation were not completely consistent with field emission-induced luminescence. A subsequent study by our group[7], which compared the spectra of radiation emitted by as-purchased and chemically-cut (and functionalized) CNTs during microwave irradiation showed that the chemically-processed CNTs produced radiation of greater intensity than the as-purchased CNTs. Well-defined photopeaks, which were present in the spectrum produced by the initial irradiation cycle, were not present in spectra produced during subsequent irradiation cycles, suggesting that they were the result of gases absorbed by the CNTs. However, there were also several other photopeaks present in the spectra produced in both initial and all subsequent irradiation cycles which were much broader and seem to be the result of some unexplained mechanism (a phenomenon which has also been noted in previous studies by other groups exploring the potential of CNT filaments[13]).

Previous theoretical studies[14,15] have suggested that long CNTs absorb microwave and radio-frequency radiation more readily than shorter CNTs. Furthermore, microwave absorption has been shown to increase as the number of CNT defects increases[16]. If the mechanism responsible for the emission of radiation by CNTs exposed to microwaves is Joule heating (which is possible, as structural imperfections are known to result in the decay of ballistic transport), one would expect to observe marked differences in the spectra of the radiation emitted by long and short CNTs, as the defect densities and absorption properties of the two types of CNTs are different. One of the primary goals of the experiments described in this report was to gain data that might help us to understand the mechanisms responsible for the emission of radiation by CNTs while being exposed to microwave fields by comparing the spectra of radiation emitted by long and short CNTs.

The second primary goal of the experiments described here was to determine how microwave irradiation, at low pressure and in the absence of chemical reagents, affects the defect densities of CNTs. This was achieved by determining the intensity ratios of the first-order Raman tangential modes (G-bands), which appear in Raman spectra between 1500 cm$^{-1}$ and 1600 cm$^{-1}$, to the D-bands, which have Raman modes between 1250 cm$^{-1}$ and 1350 cm$^{-1}$, and comparing the ratios of microwave-irradiated and as-purchased CNT samples. The G-band/D-band ratio ($I_G/I_D$) is commonly used to determine the relative amounts of impurities and/or defects in CNT samples[17]. Recently, Alvarez-Zauco et al.[5] compared the band ratios of multi-walled CNTs irradiated with microwaves at low pressure for different lengths of time, and reported changes in the ratio of less than 10%. However, the authors of this report are unaware of any studies in which the Raman $I_G/I_D$ ratios of single-walled CNTs irradiated with microwaves at low pressure and in the absence of any chemical reagents have been investigated.

## 2. Materials and Experimental Methods

Two types of CNTs were used in our study. According to the specifications provided by the supplier (Cheap Tubes Inc., USA), the first type ("short" CNTs) consisted primarily of single and double-walled CNTs with lengths ranging from 0.5 μm to 2 μm. The second type ("long" CNTs) consisted primarily of single and double-walled CNTs with lengths ranging from 5 μm to 30 μm. Both types had single and double-walled CNT contents of > 90 wt%, multi-walled CNT contents of > 5 wt%, amorphous carbon contents of < 3 wt%, and ash contents of < 1.5 wt%. Both types also had similar specific surface areas (~407 $m^2$/g), bulk densities (0.14 $g/cm^3$), and true densities (2.1 $g/cm^3$).

CNT samples with masses of 150 ± 0.1 mg were irradiated with 2.46 GHz microwaves, emitted from the antenna of a magnetron operating at 1.1 kW, while in a vertically-positioned glass tube at a pressure of ~$10^{-6}$ Torr, positioned approximately 1 cm from the magnetron antenna. Data were collected using an Ocean Optics spectrometer probe, which was also positioned approximately 1 cm from the bottom of the glass tube. The orientations of the glass tube, Ocean Optics spectrometer probe, and magnetron antenna were the same in all experiments.

The radiation spectra produced by four CNT samples (two containing short CNTs and two containing long CNTs) irradiated with microwaves for 40 s were obtained. During the experiments, the spectra of the radiation emitted by the CNTs were integrated at 0.1 s intervals during the 40 s irradiation periods.

The Raman spectra of both microwave-irradiated and as-purchased CNT samples were obtained using a Horiba Jobin Yvon LabRam HR Micro-Raman Spectrometer equipped with a 633 nm-wavelength laser.

## 3. Results and Discussion

Comparisons of the spectra of the radiation emitted by the short CNT samples and the long CNT samples in two separate experiments can be seen in Fig. 1. Background radiation has been subtracted from all spectra. The results of both experiments shown in Fig. 1 clearly show that the intensity of the radiation emitted by the long CNTs is consistently much greater than that emitted by the short CNTs. On average, the radiation emitted by the long CNTs with wavelengths between 400 nm and 1000 nm was 7.85 times more intense than that emitted by the short CNTs. As noted in previous reports[6,7], the spectra are essentially broadband in nature, suggesting that the radiation emissions are not primarily the result of some chemical mechanism involving either absorbed gases or the CNTs, themselves. However, broad photopeaks (discussed previously in this report and by others[13]) also appear in the spectra produced by both the long and short CNTs shown in Fig.1. These photopeaks suggest that, in addition to Joule heating, there is a second (unknown) mechanism that is responsible for the emission of radiation by CNTs while being exposed to microwaves.

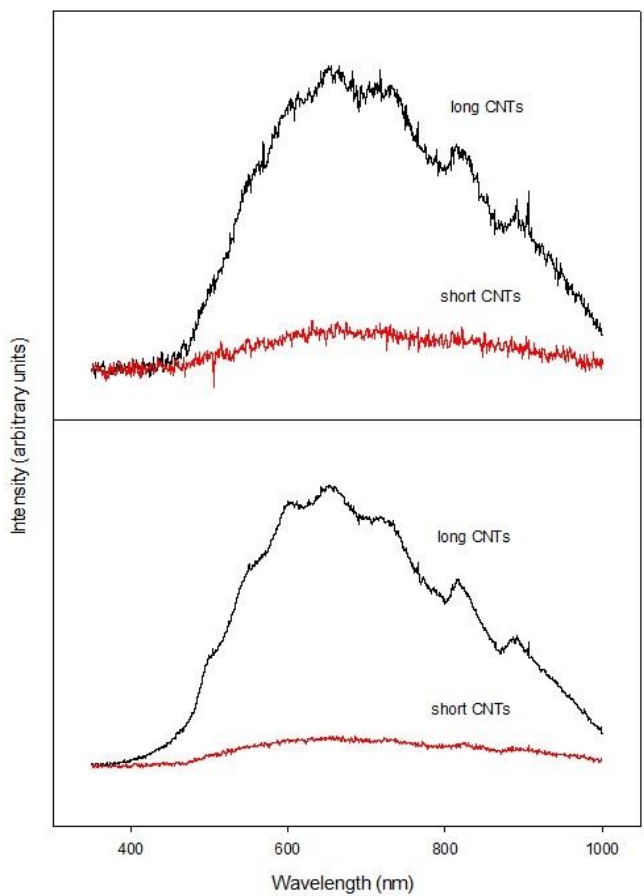

Fig. 1. Spectra of the radiation emitted from long and short CNTs during 40 s of 2.46 GHz microwave irradiation during two separate experiments

The Raman spectra of the long and short CNTs can be seen in Figs. 2 and 3, respectively. The D-modes, which appear at ~1320 cm$^{-1}$, are of low intensities and are difficult to discern in the figures. This suggests that CNT samples, as purchased, were of high purity and had low defect densities. The G-bands, which appear at ~1575 cm$^{-1}$ in Figs. 2 and 3, are much more pronounced than the D-bands.

After the subtraction of background in the spectra, the $I_G/I_D$ ratios for the as-purchased short CNTs, irradiated short CNTs, as-purchased long CNTs, and irradiated long CNTs are 26.13, 24.84, 29.64, and 17.70, respectively. As the amounts of impurities are essentially the same before and after microwave irradiation, any changes in the $I_G/I_D$ ratios would suggest changes in the defect densities of the CNT samples. In our experiments, the $I_G/I_D$ ratios decreased after microwave irradiation, suggesting that microwave irradiation at low pressure and in the absence of chemical reagents increases the defect densities of single-walled CNTs. Our results also suggest that microwave irradiation results in a greater increase in defects in the case of long single-walled CNTs, as compared to short single-walled CNTs. These results are somewhat surprising considering the results reported by Alvarez-Zauco et al.[5], which suggested that microwave irradiation of multi-walled CNTs at low pressure results in a slight (~10%) decrease in the Raman $I_D/I_G$ ratio (as opposed to the $I_G/I_D$ ratio, which decreased in the present study).

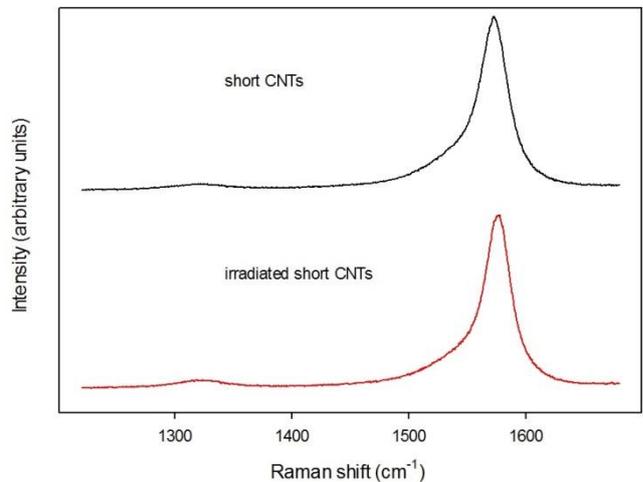

Fig. 3. Raman spectra of the as-purchased short CNTs (top) and the microwave-irradiated short CNTs (bottom)

When these results are considered in context with the results shown in Fig. 1, they strongly suggest that the primary mechanism responsible for the emission of radiation by CNTs exposed to microwaves is Joule heating. Fig. 1 shows that long CNTs emit radiation of greater intensity than short CNTs when exposed to microwaves, and the Raman $I_G/I_D$ ratios suggest that long CNTs have higher defect densities than short CNTs after exposure to microwaves, which means that one would expect a greater decay of ballistic conduction in the case of the long CNTs.

Another interesting feature is present in the Raman spectra. After irradiation, the G-bands in both the spectra produced by the short and long CNTs shifted to slightly higher wavenumbers. In the case of the as-purchased and microwave-irradiated short CNTs, the G-bands appear in the Raman spectra at 1572 cm$^{-1}$ and 1578 cm$^{-1}$, respectively; and in the case of the as-purchased and microwave-irradiated long CNTs, the G-bands appear in the Raman spectra at 1570 cm$^{-1}$ and 1580 cm$^{-1}$, respectively. Although the exact mechanism responsible for the shifts in the G-bands is unknown, previous reports[18,19] have shown that CNT strain results in G-band shifts.

The results of the present study and previous experiments[6,7,9] are in agreement with theoretical studies[14,15] that suggest that long CNTs absorb microwaves more readily than short CNTs. Our data suggest that microwave absorption results in the Joule heating of the CNTs, which leads to the emission of blackbody radiation. Theoretical treatments[20,21] have also suggested that the thermal conductivity of CNTs increases as the lengths of CNTs increase (especially for the range of CNT lengths related to the present study). The difference in thermal conductivities is primarily attributed to differences in relaxation times of N-processes[20] (in which phonon momentum is conserved, due to the sum of the phonon wave-vectors being inside of the first Brillouin zone). An increased thermal conductivity, coupled with a higher microwave-absorption cross section, would lead to long CNTs heating to higher temperatures than short CNTs under the same circumstances during microwave irradiation, which would lead to emission of blackbody radiation of much greater intensity. If the spectra shown in Fig. 1 are treated strictly as blackbody spectra, they suggest that the CNTs were heated to ~4500 K, which is consistent with previous reports where the CNT temperatures were measured by more direct methods[4,22]. Furthermore, our results suggest that microwave irradiation increases the defect densities of CNTs, which results in greater Joule heating and more intense blackbody radiation (especially in the case of long CNTs). This hypothesis would explain why previous reports[6,7] from our group have noted that the intensity of radiation emitted by the CNTs increased each time the CNTs were irradiated with microwaves, as each irradiation cycle would introduce new defects into the CNTs which would cause the decay of ballistic transport and result in both greater Joule heating and the emission of more intense blackbody radiation.

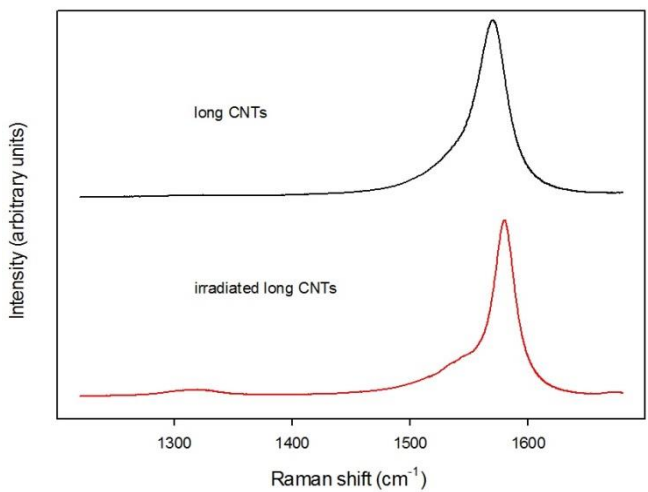

Fig. 2. Raman spectra of the as-purchased long CNTs (top) and the microwave-irradiated long CNTs (bottom)

## 4. Conclusions

In conclusion, structural modification of the CNTs due to microwave irradiation was studied by comparing Raman $I_G/I_D$ ratios, which suggests that microwave irradiation at low pressure results in an increase in CNT defects. While the $I_G/I_D$ ratios suggest that the defect density of the short CNTs increased only slightly after microwave irradiation, the data suggests that the defect density of the long CNTs increased substantially due to irradiation. The spectra of the radiation emitted by long and short single and double-walled CNTs during microwave irradiation at ~$10^{-6}$ Torr were compared. The intensity of the radiation emitted by the long CNTs was much greater than the radiation emitted by the short CNTs, which is consistent with Joule heating.

In the future, experiments should be performed to investigate the effects of microwave irradiation on individual, isolated CNTs which might provide information on the behavior of CNTs in microwaves fields without the interference of the effects of neighboring CNTs.

## Acknowledgements

The authors thank Winston Layne at the Alan G. MacDiarmid NanoTech Institute at the University of Texas at Dallas for his assistance with the Raman spectroscopy portion of this study.